\documentclass{raa}
\usepackage{graphicx,url,longtable} 

\begin{document}

\title{Non-Sinusoidal Transit Timing Variations for the Exoplanet HAT-P-12b}
\volnopage{Vol.0 (200x) No.0, 000--000}
\author{Devesh P. Sariya, \inst{1}
Ing-Guey Jiang, \inst{1}
Li-Hsin Su, \inst{1}
Li-Chin Yeh, \inst{2}
Tze-En Chang, \inst{1}
V. V. Moskvin, \inst{3}
A. A. Shlyapnikov, \inst{3}
V. Ignatov, \inst{3}
David Mkrtichian, \inst{4}
Evgeny Griv, \inst{5}
Vineet Kumar Mannaday, \inst{6}
Parijat Thakur, \inst{6}
D. K. Sahu, \inst{7}
Swadesh Chand, \inst{6}
D. Bisht, \inst{8}
Zhao Sun, \inst{9}
\and Jianghui Ji \inst{9}
}

\institute{Department of Physics and Institute of Astronomy, National Tsing-Hua University, Hsin-Chu, Taiwan; \\
\and Institute of Computational and Modeling Science, National Tsing-Hua University, Hsin-Chu, Taiwan; \\
\and Crimean Astrophysical Observatory, 298409, Nauchny, Crimea; \\
\and National Astronomical Research Institute of Thailand (NARIT), 
Siripanich Building, 191 Huaykaew Road, Muang District, Chiangmai, Thailand; \\
\and Department of Physics, Ben-Gurion University, Beer-Sheva 84105, Israel; \\
\and Department of Pure \& Applied Physics, Guru Ghasidas Vishwavidyalaya 
(A Central University), Bilaspur (C.G.) – 495 009, India; \\
\and Indian Institute of Astrophysics, Bangalore – 560 034, India; \\
\and Key Laboratory for Researches in Galaxies and Cosmology, 
University of Science and Technology of China, Chinese Academy of Sciences; \\ 
\and Purple Mountain Observatory, Chinese Academy of Sciences, Nanjing 210008\\
{\it corresponding e-mail: jiang@phys.nthu.edu.tw} \\
}

\abstract
{Considering the importance of investigating the transit timing variations
(TTVs) of transiting exoplanets, we present a follow-up study
of HAT-P-12b.
We include six new light curves observed between 2011 and 2015
from three different observatories,
in association with 25 light curves taken
from the published literature.
The sample of the data used, thus covers
a time span of $\sim10.2$ years with a large coverage of epochs (1160)
for the transiting events of the exoplanet HAT-P-12b.
The light curves are used to determine
the orbital parameters and conduct an investigation of possible
transit timing variations.
The new linear ephemeris shows a large value of reduced $\chi^2$,
i.e. $\chi^2_{red}(23)$ = 7.93, 
and the sinusoidal fitting
using the prominent frequency coming from a periodogram
shows a reduced $\chi^2$ around 4.
Based on these values and the corresponding $O-C$ diagrams,
we suspect the presence of a possible non-sinusoidal
TTV in this planetary system.
Finally, we find that a scenario with an additional
non-transiting exoplanet could explain
this TTV with an even smaller reduced $\chi^2$ value of around 2.
}

\authorrunning{Devesh P. Sariya et al.} 
\titlerunning{A follow-up transit study of HAT-P-12b} 
\maketitle

\section{Introduction}
Many generations of astronomers 
have been seraching for
the possible existence of
exoplanets before the end of twentieth century
(Briot \& Schneider 2018)
with the discovery of 51 Peg b (Mayor \& Queloz 1995).
Since then, exoplanetary science has witnessed 
a boom that has made it 
one of the most studied branches of astronomy.
Although the initial success in discovering exoplanets came from
the results of the Doppler method, the transit method has
played the most dominating role in discovering new exoplanets.
This is due to space-based surveys like
Kepler (Borucki et al 2010), K2 (Howell et al. 2014), 
Convection, Rotation and planetary Transits
(CoRoT, Baglin et al. 2006) and also, the recently launched 
Transiting Exoplanet Survey Satellite
(TESS, Ricker et al. 2015).
However, transit surveys from ground-based observing facilities
have also contributed significantly in discovering transiting exoplanets
owing to the surveys such as
the Hungarian-made Automated Telescope Network (HATNet, Bakos et al. 2004), 
The Hungarian-made Automated Telescope Network-South (HATSouth, Bakos et al. 2013), 
Trans-Atlantic Exoplanet Survey (TrES, Alonso et al. 2004), 
Super Wide Angle Search for Planets (SuperWASP, Pollacco et al. 2006), 
Kilodegree Extremely Little Telescope (KELT, Pepper et al. 2007), 
Qatar Exoplanet Survey (QES, Alsubai et al. 2013),
Multi-site All-Sky CAmeRA (MASCARA, Talens et al. 2017) survey
to name a few.
These surveys preferably detect short period, close-in planets.
Ground-based telescopes also
provide follow-up observations to confirm the transiting nature 
of exoplanets discovered from space surveys. 

The additional contributions from ground-based telescopes
are to cover a large field-of-view and to carry out
extensive follow-up observation programs which improve the
orbital parameters of a planetary system.
These observations can also be used 
for the analysis of
the transit timing variations (TTVs) 
over a longer time baseline.
In essence, a TTV is the 
transit time deviation from a linear ephemeris
which provides clues about 
the existence of another planet in the system
(Agol et al. 2005, Agol \& Fabrycky 2017, 
Sun et al. 2017, Linial et al. 2018,
Ba\c{s}t\"urk et al. 2019). 
In fact, TTVs have not only led to the discoveries of new exoplanets 
(Nesvorn\'y et al. 2012, Ioannidis et al. 2014, 
Fox \& Wiegert 2019, Sun et al. 2019), 
but also become a tool to characterize the bulk composition of 
exoplanets (Jontof-Hutter et al. 2015, Kipping et al. 2019).
Motivated by these important results, 
Holczer et al. (2016) constructed a transit timing catalog of 2599 
Kepler Objects of Interest, which will be very useful for further TTV studies.

As discussed by Ba\c{s}t\"urk et al. (2019), Saturn mass planets
are interesting for their densities and orbital properties.
HAT-P-12b is a low-density, moderately irradiated,
sub-Saturn mass ($m_p= 0.211\pm0.012 M_J$) 
transiting exoplanet whose discovery was reported by
Hartman et al. (2009) using the HAT-5 telescope (located in Arizona)
of HATNet (Bakos et al. 2004). 
HAT-P-12b orbits a moderately bright ($V\sim12.8$), 
metal-poor K4 dwarf within a period of $\sim$3.21 days
(Hartman et al. 2009). By the time of its discovery, it was 
the least massive H/He-dominated gas giant planet.
Because the above characteristics are
very different from Jupiter-mass close-in exoplanets,
HAT-P-12b has been studied through the methods of 
photometric transit observations,  
radial velocity measurements, and transmission spectroscopy
by many groups.
Lee et al. (2012) used follow-up observations to 
improve the ephemeris of the system.
Likewise, Sada et al. (2012) published $J$-band transit light curves 
for HAT-P-12b.
Todorov et al. (2013) observed the secondary eclipses using the 
IRAC instrument
on the {\it Spitzer Space Telescope}. They did not detect eclipses
at either 4.5 $\mu$m or 3.6 $\mu$m wavelengths.
The radial velocity measurements of the planet were 
produced by Kunston et al. (2014) and Ment et al. (2018). 
Mancini et al. (2018) used HARPS-N high-precision radial velocity measurements
to analyze the Rossiter-McLaughlin effect. They determined the
sky-projected obliquity ($\lambda=-54^{\circ}\,^{+41^{\circ}}_{-13^{\circ}}$)
for HAT-P-12b.
Sada \& Ramon-Fox (2016) combined publicly available light curves with radial
velocity measurements and determine physical and orbital parameters 
for HAT-P-12b.  
Spectroscopically, 
Line et al. (2013) presented an NIR transmission spectrum for the system 
using HST WFC-3. 
They found a lack of water absorption feature for 
a hydrogen-dominated atmosphere.
Alexoudi et al. (2018) performed a homogeneous analysis 
which included published data from Sing et al. (2016) and their own data
and obtained a transmission spectrum with a low-amplitude spectral slope.

The above discussion shows that
the determination of updated orbital parameters is very important
as small deviations in these values could lead to different 
physical parameters and structures of exoplanets.
Motivated by this, through a homogeneous 
long baseline TTV analysis, here
we present a comprehensive study of HAT-P-12b
with observations combining our new 
observations with the publicly available published light curves.
We include the light curves from the discovery paper 
of HAT-P-12b (Hartman et al. 2009), up to very recent 
observations, in order to cover a large range of 1160 epochs, where the   
entire data has a time baseline of $\sim10.2$ years.

The data used in this work and its basic reduction procedure
are given in Section~\ref{OBS}.
The analysis of the light curves using 
Markov-Chain Monte-Carlo (MCMC) techniques is 
described in Section~\ref{TAP}.
Section~\ref{TTV} describes a new ephemeris using linear fitting
as well as a frequency analysis and corresponding $O-C$ diagrams.
Section~\ref{dynamics} presents the dynamical two-planet model.
Finally, the conclusions of this study are presented in Section~\ref{CONC}.

\section{The Data and Reduction Procedures}
\label{OBS}

\begin{table*}
\caption{The observational log of the new data used in this work.}
\small
\centering
\begin{tabular}{ccccccc}
\hline\hline
Run & UT Date  & Instrument & Filter & Interval              & Exposure & Number of \\
    &          &            &        & ($BJD_{TDB}-2450000$) & (s)      & images    \\
\hline
1 & 2011 March 29 & Tenagra & $R$ & 5649.712644-5649.930803 & 75 & 192 \\
2 & 2011 April 17 & PM0 & Sloan $r$ & 5668.988182-5669.165950  & 50 & 159 \\
3 & 2011 April 27 & Tenagra & $R$ & 5678.633962-5678.848666 & 75 & 144 \\
4 & 2014 March 19 & P60 & $R$ & 6735.744668-6735.888703 & 15 & 276 \\
5 & 2014 April 17 & P60 & $R$ & 6764.676953-6764.805016 & 20 & 216 \\
6 & 2015 July 14  & P60 & $R$ & 7217.695978-7217.850202 & 24 & 186 \\
\hline
\label{log}
\end{tabular}
\end{table*}

Among our data, we used three transit observations 
from the 60 inch telescope (P60) installed at the
Palomar Observatory in California, USA. 
Two light curves were observed with 32 inch 
telescope at Tenagra Observatory in Arizona, USA.
The Purple Mountain Observatory's
40 inch Near-Earth Object Survey Telescope at the Xuyi Station
provided another light curve's data used in this study.
The log of the observations is listed in Table~\ref{log}.
The `run' in the table is in accordance to the date of observation.

The observed CCD images first went through some
standard procedures such as bias subtraction, flat-fielding, 
dark frames (when needed) and cosmic
rays removal with Image Reduction and Analysis Facility
(IRAF\footnote{IRAF is distributed by the National Optical 
Astronomical Observatory which is operated by the Association 
of Universities for Research in Astronomy, under contact
with the National Science Foundation.}). 
Before conducting photometry of the images, the images were first aligned 
using the `xregister' task of IRAF.
The photometry of the `cleaned' images is conducted using 
the `apphot' task in the `digiphot' routine.
The initial step in aperture photometry is to find/detect the stars in the image. 
IRAF task `daofind' finds stars in the image and lists them in a file. 
The next step in aperture photometry gives the flux value of the stars. 
IRAF task `phot' serves this purpose. 

Once we have the fluxes of stars, we conduct differential photometry.
In differential photometry, 
the target star's flux (or magnitude) is presented
with respect to one or multiple comparison stars
(e.g. Sariya et al. 2013, Jiang et al. 2013, 2016).
The selected comparison stars should not be of a variable nature.
Differential photometry cancels out the corrections required for 
the airmass and exposure time. 
It is also useful when the observing conditions are not the best.
For the HAT-P-12b data, we selected the comparison stars having the 
same instrumental magnitude and neighboring position
to the target star (HAT-P-12) in the CCD frames. 

For the TTV analysis, it is always best to include the
published light curves with the new observations
as a longer time baseline assures a better ephemeris.
We have, therefore, used three light curves from
Hartman et al. (2009), three light curves from Lee et al. (2012),
ten light curves from Mancini et al. (2018)
and nine light curves from Alexoudi et al. (2018).
The total time duration covered by the data thus becomes
slightly more than a decade.

We did not simply use the mid-transit times
for the published light curves given in the respective papers.
Instead, we applied the same procedure on those light curves 
that we applied to our data. This approach removes any systematics 
while performing parameter fitting and provides more consistent 
inputs for the TTV analysis.

The light curves were then processed through a normalization routine
to get rid of the effects caused by the airmass. 
For this purpose, we adopted the procedure described by
Murgas et al. (2014) wherein a third degree polynomial is used 
to model the airmass.
The observed flux of a light curve $F_0(t)$ can be represented as:

\begin{equation}
F_0(t)=F(t)\mathcal{P}(t),
\end{equation}
where $F(t)$ is the normalized flux of the light curve
which will be used in the further analysis
and $\mathcal{P}(t) = a_0 + a_1 t + a_2 t^2+ a_3 t^3$
is a third degree polynomial. 
A python code is used to numerically calculate the best values of
the parameters $a_0$, $a_1$, $a_2$, and $a_3$ so that the
out-of-transit part of $F(t)$ is close to unity. 

As for the timing scheme for the light curves,
we took the time from the headers of the individual images.
To make sure that the criterion used for the time is uniform, 
first we calculated observation time for the mid-exposure for every image. 
Further, it is essential to bring all the mid-exposure times to a 
common time stamp for a consistent fitting.
Hence, all the individual times of observations were converted to
the Barycentric Julian Date in 
Barycentric Dynamical Time (BJD$_{TDB}$)
following Eastman et al. (2010).

\section{The Light-Curve Analysis}
\label{TAP}

\begin{table*}
\caption{The settings of initial values and conditions for running the TAP.
The values of  $P$, $i$, 
$a/{R_*}$ and $R_{p}/R_{*}$
are taken from Hartman et al. (2009).
The value of eccentricity  ($e$) is taken from Knutson et al. (2014).
}
\centering
\begin{tabular}{ccc}
\hline\hline
Parameter & Initial Value & Condition during MCMC Chains \\
\hline
period ($P$,day)                           & 3.2130598        & Gaussian penalty with $\sigma = 0.0000021$\\
orbital inclination ($i$, $^{\circ}$) & 89.0             & Gaussian penalty with $\sigma = 0.4$ \\
scaled semi-major axis ($a/R_{*}$)         & 11.77            & free \\
planet to star radius ratio ($R_{p}/R_{*}$) & 0.1406           & free \\
mid-transit time ($T_m$)                   & TAP calculations & free, linked only for the same transit event\\
linear limb darkening ($u_1$)              & Table~\ref{LD}   & Gaussian penalty with $\sigma = 0.05$ \\
quadratic limb darkening ($u_2$)           & Table~\ref{LD}   & Gaussian penalty with $\sigma = 0.05$ \\
orbital eccentricity ($e$)            & 0.026            & Gaussian penalty with $\sigma = 0.022$ \\
longitude of periastron ($\varpi,^{\circ}$)& 0.0              & locked\\
\hline
\label{input}
\end{tabular}
\end{table*}

The transit light curves (6 new+25 published) were analyzed using
the Transit Analysis Package (TAP, Gazak et al. 2012).
TAP has previously been used by our group for 
TrES-3b (Jiang et al. 2013, Mannaday et al. (2020),
WASP-43b (Jiang et al. 2016) 
and Qatar-1b (Su et al., {\it submitted}).
TAP is an IDL based graphical user-interface driven software package
which employs the MCMC approach to 
fit the light curves using the analytic model given by Mandel \& Agol (2002)
and wavelet-based likelihood function by Carter \& Winn (2009). 

TAP involves a set of nine parameters that the user has to input.
These parameters are: 
orbital period of the planet ($P$), 
orbital inclination 
on the sky plane ($i$), 
scaled semi-major axis ($a/R_{\ast}$), 
the planet-to-star radius ratio ($R_{\rm p}/R_{\ast}$),
the mid-transit time ($T_{\rm m}$), 
the linear limb darkening coefficient ($u_1$), 
the quadratic limb darkening coefficient ($u_2$), 
orbital eccentricity ($e$) 
and
the longitude of periastron ($\varpi$).
For the input parameters mentioned above, one has to 
define one of the three conditions while running the MCMC chain of TAP.
According to the conditions, a parameter can be one of the following:
(1) completely free
(2) completely locked, or
(3) varying according to a Gaussian function.

As discussed in Section~\ref{OBS}, we dropped
the first publicly available light curve from
Hartman et al. (2009). So, the epoch zero in this study was defined by the 
second of the four publicly available light curves 
from Hartman et al. (2009).
In order to define the initial input values,
we considered most of the values mentioned in Hartman et al. (2009),
as their paper presents the maximum number of required input parameters 
and it is better for the consistency to use
input parameters from the same source.
For the eccentricity, we considered the initial input value from 
Knutson et al. (2014).
The orbital period ($P$) was defined as 3.2130598
with a Gaussian penalty of 0.0000021.
The scaled semi-major axis ($a/R_{\ast}$)
and the planet-to-star radius ratio ($R_{\rm p}/R_{\ast}$)
were chosen to be completely free and their input values were
11.77 and 0.1406, respectively. 
We also allowed the mid-transit time ($T_{\rm m}$) to be completely free
and did not input any value for it.
The longitude of periastron ($\varpi$) 
was set to 0$^\circ$ and was completely locked.
Also set with a Gaussian penalty, 
the orbital inclination on the sky plane was 
set as 89$^\circ$ with a sigma of 0.4$^\circ$.
The value of eccentricity is listed as 0.026$^{+0.026}_{-0.018}$
by Knutson et al. (2014), where we 
input the value ($e$=0.026) as a Gaussian with a sigma of 0.022, 
where the sigma
was calculated by taking the mean of errors in positive and negative directions.
The values of limb darkening coefficients were chosen
to be Gaussian with a sigma ($\sigma$) value of 0.05.
Table~\ref{input} contains the information about the input parameters 
and the condition chosen for them while running the MCMC chains.

\begin{table*}
\caption{The values of quadratic limb darkening coefficients}
\centering
\begin{tabular}{ccc}
\hline\hline
Filter & $u_1$  & $u_2$ \\
\hline
  $B$          & 0.93774724   & -0.083432883 \\
  $R$          & 0.57122572   &  0.14770584  \\
  $I$          & 0.44099208   &  0.18460748  \\
 sloan $g$     & 0.86437516   & -0.029412285 \\
 sloan $r$     & 0.60995392   &  0.13478272  \\
 sloan $i$     & 0.47080896   &  0.17786368  \\
 sloan $z$     & 0.38368084   &  0.19694884  \\
 Str\"omgren $u$ & 1.2500255    & -0.37429263  \\
\hline
\label{LD}
\end{tabular}
\end{table*}

The limb darkening is a filter dependent quantity.
All of our new light curves are in the 
Cousin $R$ band, except one light curve being in the 
Sloan $r$ band. However, the light curves we use from 
the published literature come from various filters.
These filters include Johnson $B$, Cousin $RI$,
Sloan $griz$, Gunn $gr$ and Str\"omgren $u$ band.
But the issue is that the published papers 
do not always provide the
numerical values of limb darkening coefficients they used.
Because we want to determine the mid-transit time
values using TAP instead of directly taking 
them from the concerned papers,
we decided to calculate the limb darkening coefficients
even for the published light curves.
We used the EXOFAST routine (Eastman et al. 2013)
which incorporates the quadratic limb darkening tables of 
Claret \& Bloemen (2011). 
This tool requires
some input values which were picked from 
Hartman et al. (2009) as :
effective temperature ($T_{\rm eff}$) = 4650 K, 
surface gravity (log$g$) = 4.61 ${\rm cm/s^{2} }$ and
metallicity $[Fe/H] = -0.29$.
The values of the resulting limb darkening coefficients are
listed in Table~\ref{LD}. 
As mentioned previously,
these values were defined with a Gaussian penalty
and a $\sigma$ of 0.05 while running the MCMC chains.  
EXOFAST did not output the values of
limb darkening coefficients for the Gunn-$g$ and Gunn-$r$ bands.
So, for these filters, 
we used the limb darkening coefficients 
obtained for the Sloan-$g$ and Sloan-$r$ bands instead.

One can also choose a parameter to be `linked'
among different light curves if it is not completely locked.
In the present study, we have some light curves that represent the
same transit event, and hence, the same epoch. 
We have linked the light curves representing the same epoch
together while calculating the mid-transit time for 
such light curves. If the filters were different for those 
light curves, we defined the values of limb darkening coefficients accordingly.

For each individual TAP run, five MCMC chains were calculated
and were added together to provide the final results. 
The results from TAP for the mid-transit times are given in 
Table~\ref{TAPresults}. 
Please remember that all the light curves
corresponding to the same transit event are represented by a single epoch
in the table.
Epoch numbers 346, 446 and 1144 represent multiple light curves
(see Table~\ref{TAPresults} for more information).
Owing to this reason, Table~\ref{TAPresults} contains 
25 epochs for the 31 light curves we have used.
The errors in the mid-transit time determined in this study 
for the published light curves
are consistent with the 
errors mentioned in Mallonn et al. (2015)
and Alexoudi et al. (2018) for the common light curves.
We also present the results for the photometric parameters
$a/R_{\ast}$
and $R_{\rm p}/R_{\ast}$
in Table~\ref{TAPresults} for individual epochs.
These parameters are also in agreement with the literature values.
Using radial velocity observations, 
Knutson et al. (2014) listed the value of planet's mass $m_P$,
where they mention to have used 
the sky-plane inclination $i$ from Hartman et al. (2009).
Using those, we calculated the corresponding value of $m_P {\rm sin} i$ 
for Knutson et al. (2014).
Using this $m_P {\rm sin} i$ 
and our TAP outputs for inclination during TAP runs,
we obtained the results of planet's mass according to our analysis.
Table~\ref{TAPresults} contains the TAP results for
eccentricity, inclination and planet's mass.

\begin{table*}
\centering
\caption{The results obtained from the TAP for mid-transit times 
and some photometric parameters for individual light curves.
The calculated values of planet's mass are also presented here.
Epoch here is the sequential number of transit
with respect to the reference transit light curve from Hartman et al. (2009).
The notations for the data source imply:
(a)-- Hartman et al. (2009);
(b)-- Mancini et al. (2018);
(c)-- Lee et al. (2012);
(d)-- the present work;
and 
(e)-- Alexoudi et al. (2018).
Among the data sources noted with asterisks, 
the epoch number 346 represents 4 light curves from Mancini et al. (2018).
Also, 3 light curves from 
Alexoudi et al. (2018) were observed during the same night
and thus, they belong to the common epoch number 1144. 
One of our light curves from Tenagra observatory happens
to be observed on the same night as a published light curve.
}
\vspace{0.4cm}
\tiny
\begin{tabular}{cccccccc}
\hline\hline
 Epoch & Data Source & $T_m$($BJD_{TDB}-2450000$) & $a$/$R_{\ast}$ &  $R_{\rm p}$/$R_{\ast}$ & $e$ &  $i$ & $m_P$ \\
    	& & day & & & & ($^{\circ}$) & ($M_J$)\\
\hline
     0  &         (a) &  4216.77244$^{+0.00023}_{-0.00022}$ & 11.84$^{+0.15}_{-0.19}$ & 0.1400$^{+0.0017}_{-0.0017}$  & 0.029$^{+0.021}_{-0.018}$ & 89.08$^{+0.37}_{-0.37}$ & 0.20890$^{+0.01000}_{-0.00970}$   \\
   203  &         (a) &  4869.02413$^{+0.00057}_{-0.00055}$ & 11.47$^{+0.23}_{-0.26}$ & 0.1450$^{+0.0032}_{-0.0030}$  & 0.029$^{+0.020}_{-0.017}$ & 88.96$^{+0.40}_{-0.39}$ & 0.20890$^{+0.01000}_{-0.00970}$   \\
   212  &         (a) &  4897.94185$^{+0.00084}_{-0.00089}$ & 12.19$^{+0.33}_{-0.34}$ & 0.1387$^{+0.0060}_{-0.0056}$  & 0.029$^{+0.021}_{-0.017}$ & 89.04$^{+0.39}_{-0.40}$ & 0.20890$^{+0.01000}_{-0.00970}$   \\
   346  &   (b$^{*}$) &  5328.49039$^{+0.00021}_{-0.00022}$ & 11.86$^{+0.11}_{-0.12}$ & 0.1389$^{+0.0011}_{-0.0010}$  & 0.026$^{+0.011}_{-0.011}$ & 89.12$^{+0.18}_{-0.18}$ & 0.20889$^{+0.01000}_{-0.00970}$   \\
   446  &       (c,d) &  5649.79746$^{+0.00019}_{-0.00020}$ & 11.72$^{+0.13}_{-0.15}$ & 0.1406$^{+0.0014}_{-0.0014}$  & 0.027$^{+0.015}_{-0.014}$ & 89.02$^{+0.27}_{-0.26}$ & 0.20890$^{+0.01000}_{-0.00970}$   \\
   451  &         (c) &  5665.86206$^{+0.00032}_{-0.00034}$ & 11.72$^{+0.20}_{-0.23}$ & 0.1438$^{+0.0022}_{-0.0020}$  & 0.030$^{+0.021}_{-0.018}$ & 88.95$^{+0.39}_{-0.37}$ & 0.20890$^{+0.01000}_{-0.00970}$   \\
   452  &         (d) &  5669.07486$^{+0.00077}_{-0.00082}$ & 11.77$^{+0.17}_{-0.17}$ & 0.1410$^{+0.0013}_{-0.0013}$  & 0.030$^{+0.020}_{-0.018}$ & 89.11$^{+0.35}_{-0.34}$ & 0.20889$^{+0.01000}_{-0.00970}$   \\
   455  &         (d) &  5678.71382$^{+0.00041}_{-0.00041}$ & 11.85$^{+0.20}_{-0.24}$ & 0.1370$^{+0.0019}_{-0.0020}$  & 0.029$^{+0.020}_{-0.018}$ & 89.05$^{+0.38}_{-0.39}$ & 0.20890$^{+0.01000}_{-0.00970}$   \\
   460  &         (c) &  5694.78087$^{+0.00023}_{-0.00023}$ & 11.83$^{+0.13}_{-0.16}$ & 0.1406$^{+0.0016}_{-0.0017}$  & 0.029$^{+0.021}_{-0.017}$ & 89.23$^{+0.33}_{-0.32}$ & 0.20889$^{+0.01000}_{-0.00970}$   \\
   553  &         (b) &  5993.59516$^{+0.00037}_{-0.00035}$ & 11.41$^{+0.24}_{-0.28}$ & 0.1388$^{+0.0028}_{-0.0029}$  & 0.029$^{+0.021}_{-0.017}$ & 88.63$^{+0.45}_{-0.43}$ & 0.20893$^{+0.01000}_{-0.00970}$   \\
   698  &         (b) &  6459.48810$^{+0.00019}_{-0.00020}$ & 11.66$^{+0.18}_{-0.22}$ & 0.1376$^{+0.0017}_{-0.0017}$  & 0.029$^{+0.021}_{-0.017}$ & 88.90$^{+0.39}_{-0.36}$ & 0.20891$^{+0.01000}_{-0.00970}$   \\
   783  &         (b) &  6732.59758$^{+0.00015}_{-0.00015}$ & 11.80$^{+0.10}_{-0.14}$ & 0.1453$^{+0.0011}_{-0.0010}$  & 0.030$^{+0.020}_{-0.017}$ & 89.29$^{+0.31}_{-0.33}$ & 0.20888$^{+0.01000}_{-0.00970}$   \\
   784  &         (d) &  6735.81023$^{+0.00029}_{-0.00029}$ & 11.69$^{+0.22}_{-0.25}$ & 0.1375$^{+0.0031}_{-0.0029}$  & 0.029$^{+0.020}_{-0.017}$ & 88.87$^{+0.42}_{-0.39}$ & 0.20891$^{+0.01000}_{-0.00970}$   \\
   792  &         (b) &  6761.51588$^{+0.00013}_{-0.00013}$ & 11.87$^{+0.11}_{-0.14}$ & 0.1414$^{+0.0009}_{-0.0009}$  & 0.029$^{+0.020}_{-0.017}$ & 89.26$^{+0.33}_{-0.33}$ & 0.20888$^{+0.01000}_{-0.00970}$   \\
   793  &         (d) &  6764.72820$^{+0.00040}_{-0.00040}$ & 11.99$^{+0.22}_{-0.25}$ & 0.1277$^{+0.0037}_{-0.0035}$  & 0.029$^{+0.021}_{-0.018}$ & 89.00$^{+0.38}_{-0.37}$ & 0.20890$^{+0.01000}_{-0.00970}$   \\
   928  &         (b) &  7198.49137$^{+0.00026}_{-0.00025}$ & 11.76$^{+0.18}_{-0.22}$ & 0.1473$^{+0.0028}_{-0.0030}$  & 0.029$^{+0.021}_{-0.018}$ & 89.04$^{+0.36}_{-0.36}$ & 0.20890$^{+0.01000}_{-0.00970}$   \\
   934  &         (d) &  7217.76896$^{+0.00059}_{-0.00055}$ & 11.77$^{+0.15}_{-0.15}$ & 0.1404$^{+0.0012}_{-0.0012}$  & 0.029$^{+0.020}_{-0.017}$ & 88.91$^{+0.36}_{-0.31}$ & 0.20891$^{+0.01000}_{-0.00970}$   \\
  1027  &         (e) &  7516.58280$^{+0.00025}_{-0.00025}$ & 11.66$^{+0.18}_{-0.22}$ & 0.1389$^{+0.0020}_{-0.0019}$  & 0.029$^{+0.020}_{-0.018}$ & 89.00$^{+0.38}_{-0.37}$ & 0.20890$^{+0.01000}_{-0.00970}$   \\
  1045  &         (b) &  7574.41888$^{+0.00025}_{-0.00024}$ & 11.74$^{+0.18}_{-0.22}$ & 0.1406$^{+0.0016}_{-0.0016}$  & 0.029$^{+0.020}_{-0.018}$ & 88.98$^{+0.40}_{-0.37}$ & 0.20890$^{+0.01000}_{-0.00970}$   \\
  1125  &         (e) &  7831.46347$^{+0.00032}_{-0.00031}$ & 11.72$^{+0.19}_{-0.22}$ & 0.1352$^{+0.0026}_{-0.0025}$  & 0.029$^{+0.021}_{-0.018}$ & 89.10$^{+0.37}_{-0.37}$ & 0.20889$^{+0.01000}_{-0.00970}$   \\
  1126  &         (e) &  7834.67580$^{+0.00051}_{-0.00049}$ & 11.98$^{+0.25}_{-0.28}$ & 0.1259$^{+0.0036}_{-0.0038}$  & 0.029$^{+0.020}_{-0.018}$ & 89.01$^{+0.40}_{-0.40}$ & 0.20890$^{+0.01000}_{-0.00970}$   \\
  1139  &         (e) &  7876.44492$^{+0.00022}_{-0.00021}$ & 11.84$^{+0.15}_{-0.17}$ & 0.1358$^{+0.0015}_{-0.0015}$  & 0.029$^{+0.019}_{-0.018}$ & 89.22$^{+0.32}_{-0.32}$ & 0.20889$^{+0.01000}_{-0.00970}$   \\
  1144  &   (e$^{*}$) &  7892.51130$^{+0.00014}_{-0.00015}$ & 11.81$^{+0.14}_{-0.15}$ & 0.1390$^{+0.0012}_{-0.0012}$  & 0.026$^{+0.012}_{-0.012}$ & 88.91$^{+0.25}_{-0.24}$ & 0.20891$^{+0.01000}_{-0.00970}$   \\
  1149  &         (e) &  7908.57577$^{+0.00026}_{-0.00026}$ & 11.72$^{+0.23}_{-0.26}$ & 0.1337$^{+0.0020}_{-0.0020}$  & 0.030$^{+0.020}_{-0.018}$ & 88.91$^{+0.40}_{-0.39}$ & 0.20891$^{+0.01000}_{-0.00970}$   \\
  1159  &         (e) &  7940.70745$^{+0.00037}_{-0.00038}$ & 11.81$^{+0.24}_{-0.28}$ & 0.1217$^{+0.0025}_{-0.0025}$  & 0.030$^{+0.021}_{-0.018}$ & 88.94$^{+0.41}_{-0.44}$ & 0.20890$^{+0.01000}_{-0.00970}$   \\
\hline
\label{TAPresults}
\end{tabular}
\end{table*}
\begin{table*}
\caption{A sample of the photometric light curve data of this work.
The TDB here demonstrates Barycentric Dynamical Time which is originated
from the French term `Temps Dynamique Barycentrique'.}
\centering
\begin{tabular}{cccc}
\hline\hline
Run & Epoch & TDB-based BJD & Relative Flux \\
\hline
1 & 446 &  2455649.712644 & 0.998728 \\
  &     &  2455649.713767 & 0.998597 \\
  &     &  2455649.714889 & 1.001292 \\
  &     &  2455649.716012 & 0.999719 \\
 -& - & - & -  \\
\hline
2 & 452  & 2455668.988182 & 0.996761 \\
  &      & 2455668.992052 & 0.998422 \\
  &      & 2455668.993144 & 0.996509 \\
  &      & 2455668.994237 & 0.991236 \\
	-& - & - & -  \\
\hline
 -& - & - & -  \\
\hline
\label{ourdata}
\end{tabular}
\end{table*}

\begin{figure}
\centering
\includegraphics[width=\textwidth]{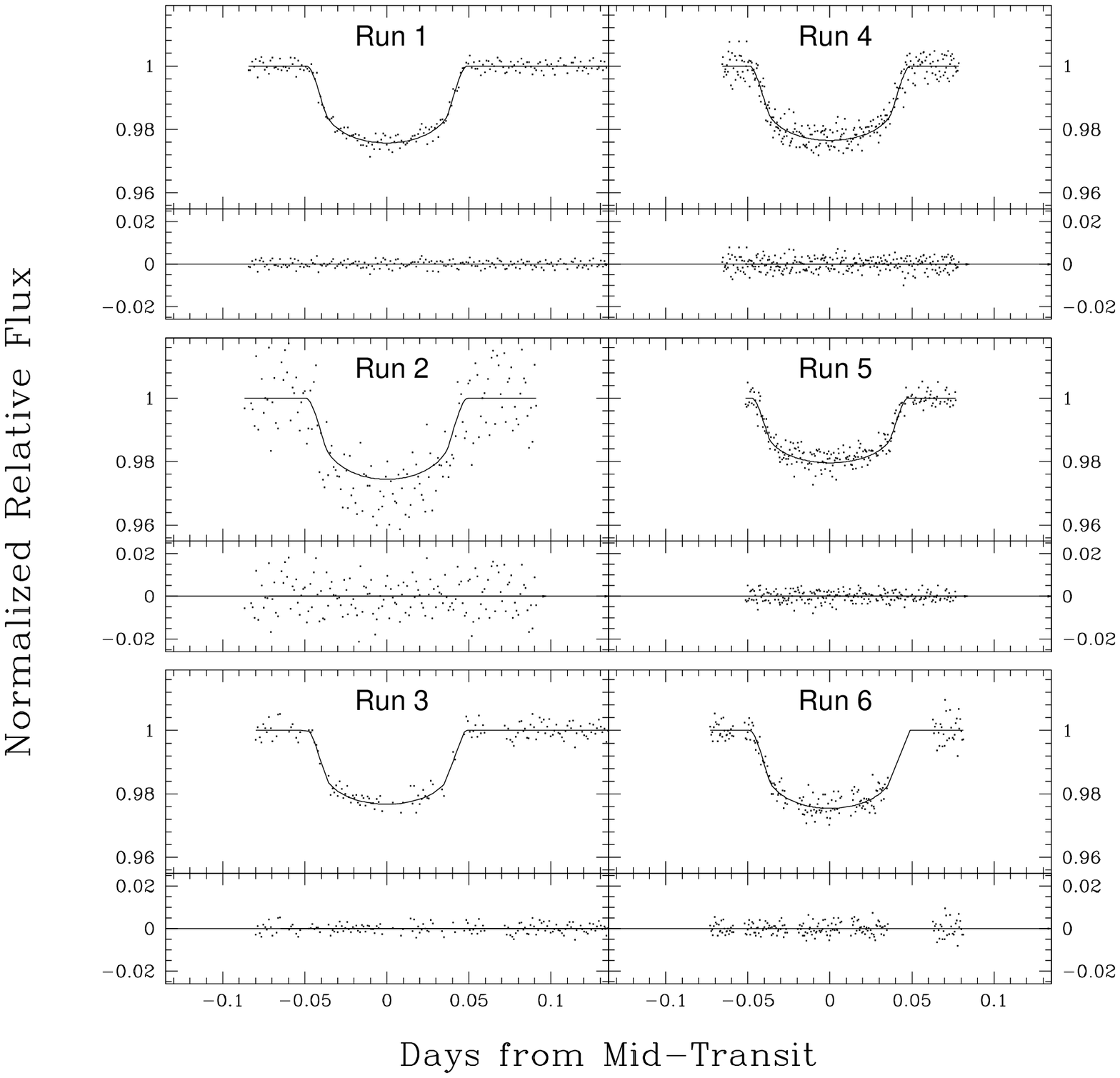}
\caption{Shown here are our observed light curves,
in the form of normalized relative flux plotted with
days from mid-transit time. The data points represent the flux and the
curves show the TAP fitting result. The corresponding residuals are shown
in the bottom panels for each `run'.} 
\label{lightcurves}
\end{figure}

Our normalized observational light curves 
and the corresponding TAP fitting 
with $x-$axis adjusted for the
mid-transit time are shown in Fig.~\ref{lightcurves}.
The light curves are in the sequence of the `run' defined in 
Table~\ref{log}. 
A few lines of our photometric observations
with $BJD_{TDB}$ and normalized relative flux are given in Table~\ref{ourdata}.
The full version of this table will be provided in
the machine readable format with this paper.

\section{The TTV Analysis}
\label{TTV}

\subsection{The Linear Fit and a New Ephemeris}

Once we have all the mid-transit times in
BJD$_{TDB}$, we can determine a new ephemeris
by $\chi^2$ minimization of the following linear relation:

\begin{equation}
T^C_{\rm m} (E)=T_0+PE,
\end{equation}

where $P$ and  $E$ are period and epoch.
The reference time $T_0$ was arbitrarily chosen to be at epoch $E=0$.
For an individual epoch $E$, $T^C_{\rm m} (E)$ is the calculated mid-transit time.
Using linear fitting, we obtain 
$T_{0} = 2454216.773311 \pm 0.000293 $  ($BJD_{TDB}$) and
$P = 3.21305762 \pm 0.00000036 $ (day).

\begin{figure}
\centering
\includegraphics[width=\textwidth]{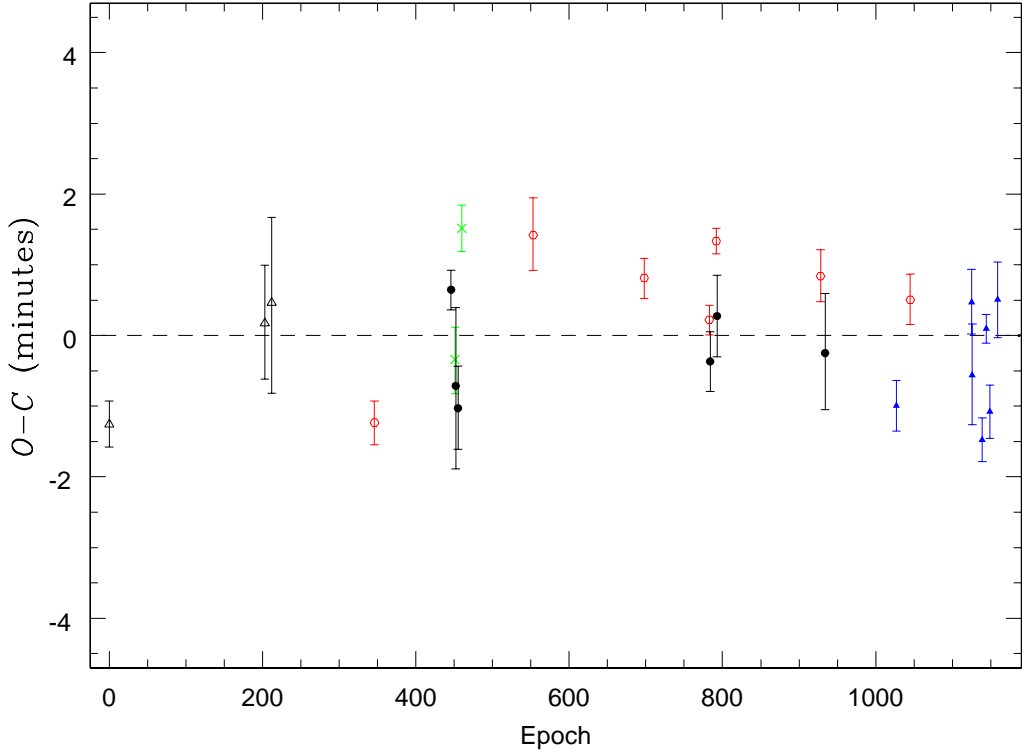}
\caption{The $O-C$ diagram for the linear fitting.
Here, the notations of the data points are as follows:
filled black circles-- our data points;
open black triangles-- Hartman et al. (2009);
green crosses -- Lee et al. (2012);
red open hexagons-- Mancini et al. (2018); and
filled blue triangles -- Alexoudi et al. (2018).
}
\label{OCLinear}
\end{figure}

If $\sigma_i$ is the mean of the error in the positive and 
negative directions of an observed mid-transit time
given by TAP, then using observed 
and calculated values of mid-transit times,
the $\chi^2$ of the fitting is determined using the formula:\\

\begin{equation}
\chi^2 =\sum_{i=1}^{N} \frac{({O}_i-{C}_i)^2}{{\sigma}_i^2},
\end{equation}

where ${O}_i$ is an observed mid-transit time,
${C}_i$ is a calculated mid-transit times, and $N$ is the 
number of included epochs. 
The value of the $\chi^2$ for the linear fitting is 
182.49.
There are 23 degrees of freedom in our model, so
the reduced $\chi^2$
becomes, 
$\chi^2_{red}(23)$ = 7.93.
This large value of the $\chi^2_{red}$ in the linear fitting 
can be
indicative of the presence of TTVs. 
Ideally, when there is no TTV, 
the time between any two adjacent transit events  
should be exactly equal to the orbital period. 
The $O-C$ diagram for the linear fitting is presented in
Fig.~\ref{OCLinear}, which shows deviation between
the observed mid-transit time and the one predicted by
a simple two-body orbit.

\subsection{The Frequency Analysis}
\label{FREQ}

\begin{figure}
\centering
\includegraphics[width=\textwidth]{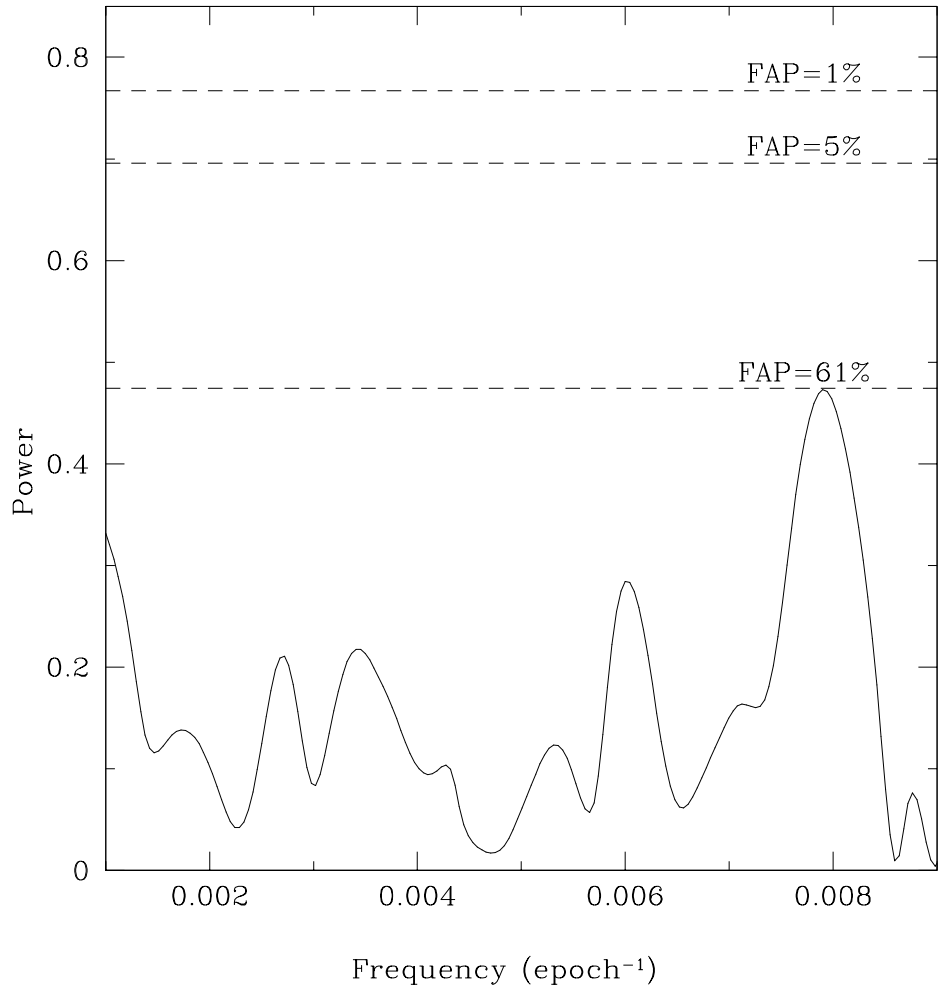}
\caption{
A periodogram showing the spectral power
versus frequencies determined for all the data (new+published)
used in this paper.
The FAP for our largest power frequency is 61\%.
In the plot, the powers corresponding to the FAP values of
1\% and 5\% are also shown with dotted lines.
}
\label{periodogram}
\end{figure}

We searched for possible frequencies which might be causing
variation in the data using 
generalized Lomb-Scargle periodogram
(Lomb 1976, Scargle 1982, Zechmeister \& Kuerster 2009).
This procedure considers the error bars while determining the periodogram.
The periodogram is shown in the Fig.~\ref{periodogram}.
If $f$ is the frequency related to the highest peak of power
in the periodogram, then the possible TTVs are tested by  $\chi^2$
minimization of the following equation that consists of
both linear and sinusoidal terms:
\begin{equation}
T_S (E) = P E+b+z\sin(2\pi f E -\phi).
\end{equation}

\begin{figure}
\centering
\includegraphics[width=\textwidth]{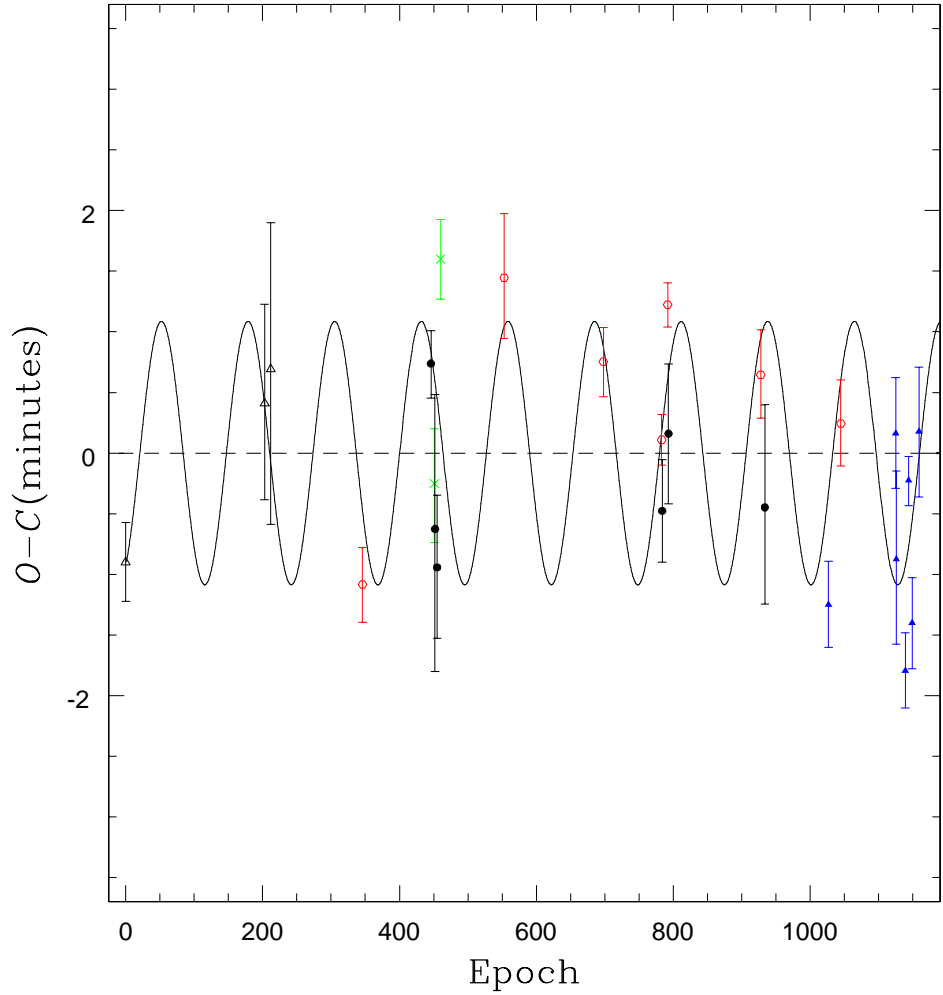}
\caption{
The $O-C$ diagram for  one-frequency model
The models is determined for
$f= 0.00790059461$ epoch$^{-1}$
The curve shows the fitting function.
Among the data points, the filled black circles represent our data;
open black triangles are for the Hartman et al. (2009) data;
Lee et al. (2012) data is shown by green crosses;
red open hexagons are for the Mancini et al. (2018) data points; and
filled blue triangles represent the data from Alexoudi et al. (2018).
}
\label{sinefitallfreq}
\end{figure}

In the equation above, the predicted mid-transit time at a given epoch
$E$ is $T_S (E)$
while $P, b,$ amplitude $z$ 
and phase $\phi_1$ are the fitting parameters.
The frequency corresponding to the highest power peak 
($f= 0.00790059461$ epoch$^{-1}$,
allows us to determine the fitting parameters:
$P= 3.21305803 \pm 0.00000019 $ day,
$b= 2454216.773065 \pm 0.000145 $ day,
$z= -0.000754 \pm 0.000080 $ day,
and 
$\phi= 4.163 \pm 0.109$ rad.
The value of the $\chi^2$ is 88.02.
For 25 data points, we are determining four 
parameters from fitting. 
This model has 21 degrees of freedom, where
the value of the reduced $\chi^2$ decreases to 4.19.
The $O-C$ diagram as a function of epoch $E$, 
for one frequency scenario is given in Fig.~\ref{sinefitallfreq}.
The $O-C$ value shown in the curve
depicts the value of $T_S (E)$ 
minus the linear term ($P E+b$). 
The data points representing the light curves 
are also adjusted according to the fitting and are shown in the figure.

The false-alarm probability (FAP) 
was determined following the procedure explained in 
Press et al. (1992). 
As can be seen in Fig.~\ref{periodogram}, the FAP 
for the frequency with maximum power is 61\%. 

Only the peaks with a significantly high signal-to-noise power ratio
should be considered from a periodogram
(Breger et al. 1993; Kuschnig et al. 1997).
Since no other peak shown in Fig.~\ref{periodogram} has a high S/N ratio,
we did not consider any other peak for frequnecy analysis. 

To conclude, considering that the  $\chi^2_{red}$ value is around 4 and
the FAP is large, the possible TTVs are probably of the non-sinusoidal nature.

\section{The Two-Planet Model}
\label{dynamics}

The values of $\chi^2_{red}$  in the above analysis indicate
possible non-sinusoidal TTVs in HAT-P-12 planetary system. 
In order to probe a physical scenario for the explanation
of these TTVs, we explore the possibility to have an additional 
exoplanet (HAT-P-12c) in this system (see, Nesvorn\'y et al. 2012 for example). 

In the approach we used, by feeding some assumed initial input values of 
the parameters for both the planets, HAT-P-12b and HAT-P-12c, 
the theoretical TTVs are produced through the dynamical calculations of 
the {\it TTVFast} code (Deck et al. 2014). 
These theoretical TTVs are used to fit our observational mid-transit times.
The best-fit model can be obtained
through a MCMC sampling code {\it MC3} (Cubillos et al. 2017). 

Before running the MCMC sampling, we first need to set
the distributions and the ranges of numerical values of photometric parameters 
for both planets.
For HAT-P-12b, the parameters are already determined in
the previous sections of this paper. 
So, these parameters were set as either fixed values or with a 
certain range around the previously determined values. 
For example, since the orbital period can vary slightly
during the orbital integration, we provide a total interval width of 
0.2 day for the orbital period of HAT-P-12b.
The orbital eccentricity and inclination of HAT-P-12b 
are taken as the mean values 
of the results shown in Table~\ref{TAPresults}.
In order to search for the best-fit model for the new exoplanet
HAT-P-12c, the initial input values were set within larger 
ranges and are set to be uniformly distributed.
Also note that the mass of central star (HAT-P-12) is set to be 
0.733 $M_\odot$ according to Hartman et al. (2009).
Table~\ref{MC3input} gives a summary of the input parameters for {\it TTVFast}.
For the parameters with a defined range of input values, 
the values are given inside the brackets, [ ]. 
\begin{table*}
\caption{The parameter setting of the two-planet model. 
The notations and units of the parameters are also given in column 1.
The range of input values for some parameters are defined in [ ]. }
\centering
\begin{tabular}{ccc}
\hline\hline
Parameter & HAT-P-12b & HAT-P-12c \\
\hline
mass ($m_{\rm p}$, $M_{\rm J}$) & [0.205, 0.213] & [0.0001, 1]       \\ 
period ($P$,day)           & [3.1130, 3.3130] &  [3.3, 16.5]      \\
orbital eccentricity ($e$) & $0.02898^{(\#)}$ & [0.0, 0.2] \\
orbital inclination ($i$,$^{\circ}$) & $89.02176^{(\#)}$ &[59.02176, 119.02176] \\
longitude of ascending node ($\Omega$, $^{\circ}$)& $0.0^{(\#)}$  & [-30, 30]   \\
argument of pericenter ($\omega$,  $^{\circ}$)   &  $0.0^{(\#)}$  & [0, 360]  \\
mean anomaly ($M$, $^{\circ}$)                   &  $(\#\#)$  & [-180, 180]   \\
\hline
\multicolumn{3}{l}
{\footnotesize Remark $(\#)$ indicates that the parameter is fixed.}\\
\multicolumn{3}{l}
{\footnotesize Remark $(\#\#)$ indicates that the mean anomaly of HAT-P-12b 
is determined by other parameters.}\\
\label{MC3input}
\end{tabular}
\end{table*}

\begin{figure}
\centering
\includegraphics[width=\textwidth]{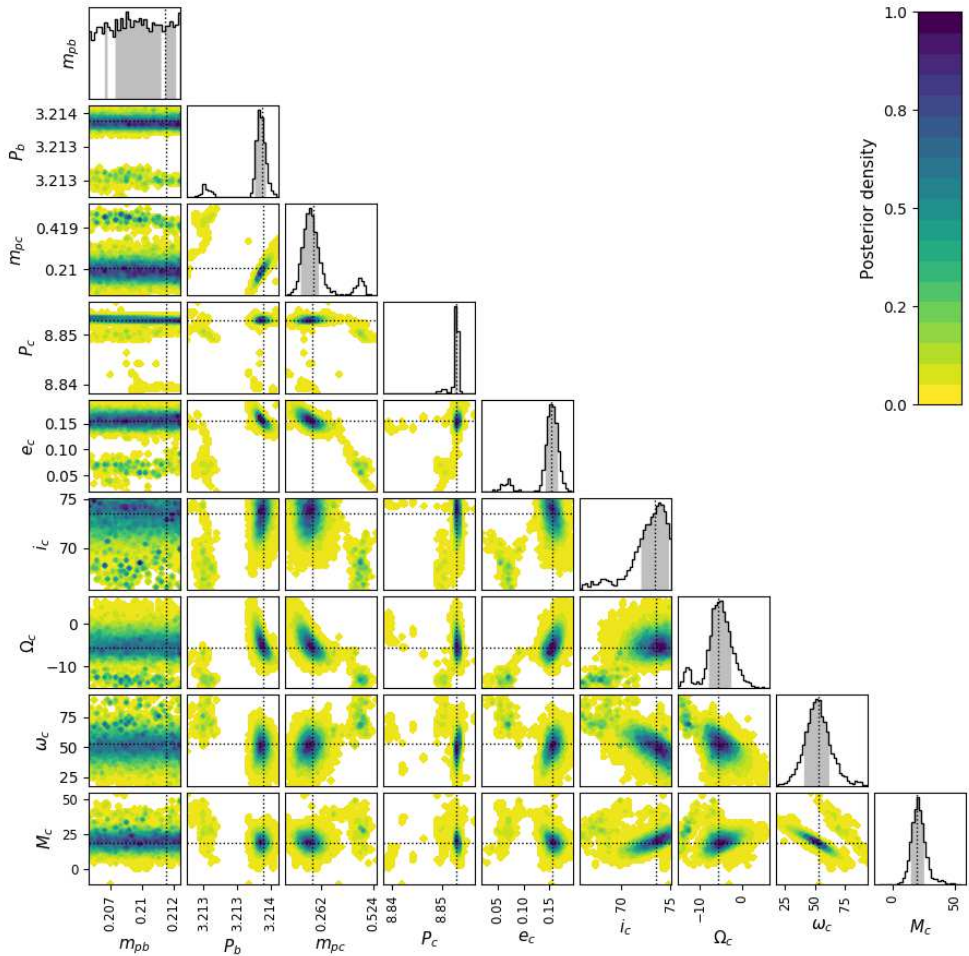}
\caption{
The MCMC posterior parameter distributions. 
Those with colors are 
pairwise two-dimensional projections. The one-dimensional projections
are presented as the histograms on the top.
The parameters shown here have one more subscript but their meanings 
and units are the same as those in Table~\ref{MC3input}.  
The parameters with subscript `$b$' are for the exoplanet HAT-P-12b,
and those with subscript `$c$' are for the exoplanet HAT-P-12c.
The dotted lines indicate the parameter 
values of the best-fit model.
The grey areas of histograms denote $68\%$ highest 
posterior density regions of the parameter distributions.
}
\label{figMCMC}
\end{figure}

As shown in Table~\ref{MC3input}, nine parameters can change 
their values during the MCMC sampling,  
with a total number of samples being $2\times 10^{7}$.
After we obtain the above result, 
in order to have more MCMC samples 
with parameters closer to the best-fit model, 
the MCMC sampling is executed again while nine parameters are now 
within smaller ranges as shown in Fig.~\ref{figMCMC}.   
Fig.~\ref{figMCMC} presents 
the MCMC posterior distributions of these nine parameters.
Both two-dimensional and one-dimensional projections
are plotted. 
Those parameters with subscript `$b$' are for the exoplanet HAT-P-12b,
and those with subscript `$c$' are for the exoplanet HAT-P-12c.  
These distributions give the probabilities that 
particular numerical values are employed during the MCMC sampling.
The color panels are the pairwise distributions.
The histograms are the one-dimensional distributions, 
where grey areas denote $68\%$ highest 
posterior density regions of the distributions.
The dotted lines in the figure 
indicate the values of the best-fit model.
The corresponding results for the parameter values 
according to this best-fit model are given in Table~\ref{bestmodel}. 
Since nine parameters are being determined, there are 
16 degrees of freedom.
The reduced $\chi^2$ of this best-fit model thus becomes 
$\chi^2_{red}(16)$ = 2.09, which is much smaller than 
the reduced $\chi^2$ values during linear fitting and frequency analysis.

\begin{figure}
\centering
\includegraphics[width=\textwidth, angle=90]{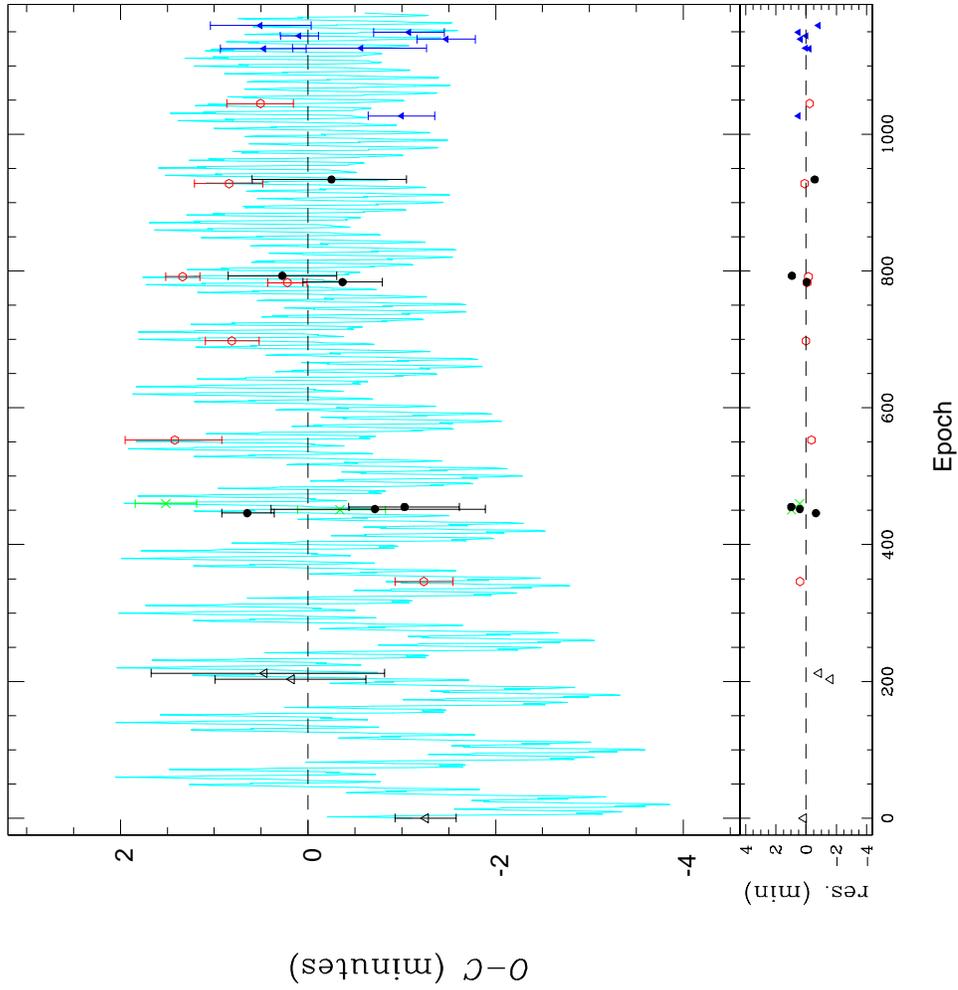}
\caption{
The $O-C$ diagram for the two-planet model.
The curve shows the theoretical TTV and the points with error bars
are the same as those shown in Fig.~\ref{OCLinear}.
The bottom panel shows a distribution of the fitting residuals
between the data points and the model.}
\label{figdynamicalTTV}
\end{figure}

The theoretical TTVs of this best-fit, two-planet model
are plotted as the curve shown in Fig.~\ref{figdynamicalTTV}. 
The data points with error bars in this figure are
$O-C$ values for the observational data (same as Fig.~\ref{OCLinear}).
The bottom panel in Fig.~\ref{figdynamicalTTV} shows 
the residuals of fitting the model to the $O-C$ data points.
The standard deviation of the residuals is 0.61 min
while the average value of the means of error bars for 
$O-C$ values is $\sim$0.49 min.
It is evident from Fig.~\ref{figdynamicalTTV} that 
the theoretical curve lies within the error bars of 
observational data for most of the epochs. 
Therefore, judging from a reasonably good data fitting
and a smaller value of the reduced $\chi^2$, 
we deduce that this two-planet model
could explain the observational TTV of HAT-P-12 planetary system.  

Considering the orbital period and inclination of HAT-P-12c 
(Table~\ref{bestmodel}),
it is impossible to have transits for this exoplanet 
unless its radius is larger than 40 times that of Jupiter. 
This explains why there are no observed transit events for this exoplanet.

\begin{table*}
\caption{Results from the best-fit model for two-planet scenario.
The subscripts $b$ or $c$ are added to distinguish between the two planets.}
\vspace{0.5mm}
\centering
\begin{tabular}{ccccccccc}
\hline\hline
$m_{\rm pb}$ & $P_{\rm b}$ & $m_{\rm pc}$ & $P_{\rm c}$ & $e_{\rm c}$ & 
$i_{\rm c}$  &  $\Omega_{\rm c}$ &  $\omega_{\rm c}$ & $M_{\rm c}$\\
($M_{\rm J})$  & (day) & ($M_{\rm J}$) & (day) &  & 
($^{\circ}$) & ($^{\circ}$) & ($^{\circ}$) & ($^{\circ}$) \\
\hline
0.212 & 3.2134 &  
0.218 & 8.8530 & 0.15499 &  73.49569 & -5.58 & 52.785 & 18.892\\ 
\hline
\label{bestmodel}
\end{tabular}
\end{table*}

\section{Conclusions}
\label{CONC}

Using the telescopes from three observatories, 
we present six new light curves of the transiting exoplanet
HAT-P-12b. These observations were combined with 25 light curves from
published papers to further enrich the baseline of data to 1160 epochs. 
A self-consistent homogeneous analysis was carried out for all
the light curves to make sure 
that our TTV results are not affected by any systematics.
The photometric parameters determined by us are in agreement to their
values in earlier published works.
We determined new ephemeris for the HAT-P-12b system
by a linear fit and sinusoidal curve fitting for different frequencies.
The values of reduced $\chi^2$ from the linear fitting is 
7.93 
while from the sine-curve fitting 
for the highest power frequency, 
the value of $\chi^2_{red}$ is obtained as 
4.19. 
These values and the large FAP indicate that the 
TTV could be non-sinusoidal. 

Finally, through a MCMC sampling, a two-planet model
is found to be able to produce a theoretical TTV 
which could explain the observations to a satisfactory level with 
a value of $\chi^2_{red}$ = 2.09. 
Therefore, a scenario 
with a non-transiting exoplanet might explain the
TTV of HAT-P-12b.

To conclude, our results show the existence of non-sinusoidal
TTVs. Though a two-planet model could lead to a better fitting, 
the validation of a new exoplanet is out of the scope and not provided here.
Hopefully, the nature of this system could be further understood
in the future.

\begin{acknowledgements}
We are thankful to the referee of this paper for very helpful suggestions.
This work is supported by the grant from
the Ministry of Science and Technology (MOST), Taiwan.
The grant numbers are
MOST 105-2119-M-007 -029 -MY3
and MOST 106-2112-M-007 -006 -MY3. 
Devesh P. Sariya is grateful to
Chiao-Yu Lee, Dr. Chien-Yo Lai, and Dr. Arti Joshi
for useful discussion on data reduction,
and also the Crimean Astrophysical Observatory (CrAO) for 
the hospitality and exchange of knowledge during his research visit.
P.T. and V.K.M. acknowledge the University Grants
Commission (UGC), New Delhi, for providing the financial
support through Major Research Project no. UGC-MRP 43-521/2014(SR).
P.T. expresses his sincere thanks to IUCCA, Pune, India for providing 
the supports through IUCCA Associateship Programme.
D. Bisht is financially supported by
the Natural Science Foundation of China
(NSFC-11590782, NSFC-11421303).
JJH is supported by the B-type Strategic Priority Program of the Chinese
Academy of Sciences (Grant No. XDB41000000), the National Natural Science
Foundation of China (Grant No. 11773081), CAS Interdisciplinary Innovation
Team, Foundation of Minor Planets of the Purple Mountain Observatory.
\end{acknowledgements}


\clearpage

\clearpage

\clearpage

\clearpage

\clearpage

\clearpage

\end{document}